\documentclass[12pt,a4paper]{article}
\usepackage{amsmath}
\usepackage[dvips]{graphicx}
\input epsf
\usepackage{times}
\begin{document}
\begin{titlepage}
\title{Semiclassical mechanism for single-spin asymmetry in $\pi^0$-production}
\author{S.M. Troshin,
 N.E. Tyurin\\[1ex]
\small  \it Institute for High Energy Physics,\\
\small  \it Protvino, Moscow Region, 142280, Russia} \normalsize
\date{}
\maketitle

\begin{abstract}
The chiral quark model combined with unitarity and impact
parameter picture provides simple semiclassical mechanism for
generation of the single-spin asymmetry $A_N$ in the $\pi^0$-production
in the polarized proton collisions at RHIC.
We  derive its linear $x_F$-dependence
in polarized proton fragmentation region
along with the  energy and transverse momentum independence at
large $p_T$ values.

\end{abstract}
\end{titlepage}
\setcounter{page}{2}

\section*{Introduction}
Single-spin asymmetry (SSA) is a sensitive tool to probe QCD at
small and large distances. Experimentally significant SSA was
observed in various processes of elastic scattering and inclusive
hadron production. Study of transverse single-spin asymmetries in
deep-inelastic processes (DIS)  observed a significant progress
during last years; it has been shown that  asymmetry can be
related to a  rescattering in the final-state interactions due to
gluon exchange \cite{brodsky,metz} -- coherent effect not
suppressed in the Bjorken limit. Another step in this direction is
the nonperturbative, instanton-induced mechanism of SSA generation
\cite{koch,diakp,shur,hoyer}. Instantons lead to
 appearance of the quark anomalous magnetic moment \cite{koch,diakp} and soft
 rescattering can provide a source for a leading twist SSA in
 semi-inclusive DIS \cite{hoyer}. Moreover, as it was noted there
 it would also affect integrated parton distributions and may jeopardize QCD
 factorization. Pauli coupling have an important consequences opening possibility
of transversity studies in DIS processes  \cite{hoyer}.
A unified picture of SSA generation in Drell-Yan processes which
combines Sivers mechanism with account for the higher twists
contributions was proposed recently in \cite{ji}.
Thus the significant role of non-perturbative effects in the mechanism of SSA
generation is becoming more and more evident nowadays.

The processes of hadron-hadron interactions are even more complicated
than DIS and origin of SSA in these reactions is not fully clarified. Despite
a great progress in theoretical studies devoted to this problem, the phenomenological
success is rather limited;  a comprehensive approach  able to get
description of the existing set of the  experimental data on polarization, asymmetries,
spin correlation parameters and cross-sections is still absent at the moment.

The most
 widespread  approaches in the field of hadronic processes are based on the
assumed extended factorization in QCD with account for the
internal parton transverse momenta in structure functions \cite{sivers,anselm,burk}
 or in the fragmentation function \cite{collins,others}. An account for the direct
higher twists contributions to parton scattering subprocesses
 also can lead to a nonzero asymmetry \cite{efrem,sterm,koike}.

 However,  as it was shown  recently the Collins frgmentation mechanism is suppressed
 in hadronic pion production \cite{anselmcol}. It was shown in this work that the Sivers
 effect gives dominating contribution to asymmetry and the second relevant contribution
 gives a convolution of transversity with Collins fragmentation while other contributions to
 asymmetry can be neglected.
It can lead to a  significant contribution to
 $A_N$ for the quark Sivers mechanism at moderate $p_T$ and to its decrease  at higher
 transverse momenta. Note, that the asymmetry is predicted to be non-zero at $x_F=0$ for
 the both quark and gluon Sivers mechanisms and the same is valid for the asymmetry $A_N$
 for the gluon Sivers mechanism at $x_F<0$.

Decreasing dependence of SSA with $p_T$ has not yet been observed experimentally, most experimental
data are consistent with a flat transverse momentum dependence at $p_T>1$ GeV/c. Another important
 point regarding unpolarized inclusive cross-section of $\pi^0$-production was discussed in
\cite{bsof}: it is unclear
are the above approaches able to describe unpolarized inclusive cross-section
 dependence on transverse momenta.
It has also been shown in the above paper that the description of the inclusive cross-section
for $\pi^0$-production, at the energies lower than the RHIC energies meets difficulties in
the framework of the perturbative QCD.
Role of higher twist
contributions was studied in the recent analysis of pQCD scaling
  of inclusive cross--section at large $p_T$ and
  its experimental status was given  in \cite{brpir}.
 Deviation from the pQCD scaling is mostly
 noticeable in the forward region where the most significant asymmetry in the $\pi^0$ production
 in $pp_\uparrow \pi^0 X$ has also been observed by the STAR collaboration at RHIC \cite{star} at
$\sqrt{s}= 200$ GeV (in the fragmentation region of the polarized proton).
At the same time $A_N=0$ in the neutral pion production
   in the backward  and midrapidity regions \cite{phenix,proza}.
   SSA has also a zero value in the $pp_\uparrow\to pX$,
   while $A_N\neq 0$ in the $pp_\uparrow\to nX$ \cite{togawa}
   in the polarized proton fragmentation region.
The  experimental features observed at RHIC represent a difficulty for the explanation
 in the above mentioned theoretical approaches, in particular, those based on Sivers mechanism
(is at variance with zero asymmetry
at $x_F\leq 0$) or
account for the anomalous magnetic moment of quarks (is at variance with zero asymmetry
 in the process $p_\uparrow p\to pX$).
Of course, more experimental data are needed to perform a conclusive test of various theoretical predictions
and those predictions should be more specified and elaborated for the observables at the hadronic level.

Keeping in mind the experimental and theoretical situation in the field of SSA studies
we show in this note that the gross features of SSA measurements at RHIC and FNAL (linear
increase of asymmetry with $x_F$ and flat transverse momentum dependence at $p_T>1$ GeV/c)
    can be explained and qualitatively described
    in the framework of the simple semiclassical mechanism based
   on  the further development
   of the specific chiral quark model \cite{csn} and results of its adaptation
   for the treatment of the polarized and unpolarized inclusive cross-sections  \cite{unpol}.
   The  data of STAR collaboration \cite{star}
    for the unpolarized inclusive cross-section  can simultaneously
     be described. This mechanism is consistent with other new experimental facts
found at RHIC.

\section{Semiclassical mechanism of SSA generation}

It might happen  that
the SSA originates from the
nonperturbative sector of QCD and is related to
the mechanism of spontaneous chiral symmetry breaking ($\chi$SB) in QCD \cite{bjorken},
 which  leads
to generation of quark masses and appearance of quark condensates. This mechanism describes
transition of current into  constituent quarks, which are
   the quasiparticles with masses
 comparable to  a hadron mass scale.   The other well known direct result of $\chi$SB
   is  appearance of the Goldstone bosons.
Thus constituent quarks and Goldstone bosons are the effective
degrees of freedom in the chiral quark model.

Thus we consider a
 hadron as an extended object consisting of the valence
constituent quarks located in the central core which is embedded into  a quark
condensate. Collective excitations of the condensate are the Goldstone bosons
and the constituent quarks interact via exchange
of Goldstone bosons; this interaction is mainly due to a pion field and of the
spin--flip nature \cite{diak}.

At the first stage of hadron interaction common effective
self-consistent field  appears. This field is generated by $\bar{Q}Q$ pairs and
pions interacting with quarks. The time of  generation of the effective field $t_{eff}$
\[
t_{eff}\ll t_{int},
\]
where $t_{int}$ is the total interaction time. This assumption on the almost instantaneous
generation of the effective field has some support in the very short thermalization time revealed
in heavy-ion collisions at RHIC \cite{therm}.

Valence constituent quarks   are
 scattered simultaneously (due to strong coupling with Goldstone bosons)
and in a quasi-independent way by this effective strong
 field. Such ideas were  used in the model \cite{csn} which has
been applied to description of elastic scattering and hadron production \cite{unpol,mult}.

In the initial state of the reaction $pp_\uparrow\to \pi^0 X$ the proton is polarized
 and can be represented in the simple SU(6) model as  following:
 \begin{equation}\label{pr}
 p_\uparrow=\frac{5}{3}U_\uparrow+\frac{1}{3}U_\downarrow+\frac{1}{3}D_\uparrow+
 \frac{2}{3}D_\downarrow.
\end{equation}
We will exploit the common feature of chiral quark models:
the constituent quark $Q_\uparrow$
with transverse spin in up-direction can fluctuate into Goldstone boson and
  another constituent quark $Q'_\downarrow$ with opposite spin direction,
   i. e. perform a spin-flip transition \cite{cheng}:
\begin{equation}\label{trans}
Q_\uparrow\to GB+Q'_\downarrow.
\end{equation}

The $\pi^0$-fluctuations of quarks do not change the quark
 flavor and assuming they have
 equal probabilities in the processes:
\begin{equation}\label{transu}
U_{\uparrow,\downarrow}\to \pi^0+U_{\downarrow,\uparrow}
\quad
\mbox{and}
\quad
D_{\uparrow,\downarrow}\to \pi^0+D_{\downarrow,\uparrow},
\end{equation}
the production of $\pi^0$ by the polarized proton $p_\uparrow$ in this simple $SU(6)$
picture can be regarded as
a result of the fluctuation of the constituent quark $Q_\uparrow$ ($Q=U$ or $D$) in the effective
field into the system $\pi^0+Q_{\downarrow}$ (Fig. 1).
\begin{figure}[h]
\begin{center}
  \resizebox{6cm}{!}{\includegraphics*{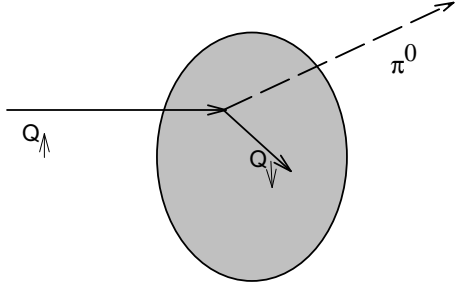}}
\end{center}
\caption{Schematical view of $\pi^0$--production in polarized proton-proton interaction.
 \label{ts1}}
\end{figure}

The contributions to the cross-sections difference
of the quarks polarized in opposite directions compensate each other (as it will be clear in
 what follows),
and it is not the case for the
$\pi^0$-production in the unpolarized case. Therefore the asymmetry $A_N$ should
obey the inequality
$|A_N(\pi^0)|\leq 1/3$.

To compensate quark spin flip $\delta {\bf S}$, an orbital angular momentum
$\delta {\bf L}=-\delta {\bf S}$ should be generated in the final state of reaction (\ref{trans}).
The presence of this orbital momentum $\delta {\bf L}$  in its turn
means  shift in the impact parameter
value of the Goldstone boson $\pi^0$:
\[
\delta {\bf S}\Rightarrow\delta {\bf L}\Rightarrow\delta\tilde{\bf b}.
\]
Due to   different strengths of interaction at the different
impact distances, i.e.
\begin{eqnarray}
\nonumber p_\uparrow\Rightarrow   Q_\uparrow & \to & \pi^0 + Q_\downarrow\Rightarrow\;
\;-\delta\tilde{\bf b}, \\
\label{spinflip}p_\downarrow\Rightarrow Q_\downarrow & \to & \pi^0 + Q_\uparrow\Rightarrow\;
\;+\delta\tilde {\bf b}.
\end{eqnarray}
the processes of transition $Q_\uparrow$ and $Q_\downarrow$ to $\pi^0$
 will have different probabilities which  leads eventually to nonzero asymmetry
$A_N(\pi^0)$.
Eqs. (\ref{spinflip}) clarify mechanism of the SSA generation:
when shift in impact
parameter is $-\delta\tilde {\bf b}$ the
interaction is stronger than when the shift is $+\delta\tilde {\bf b}$,
and the asymmetry $A_N(\pi^0)$ is positive.
It is important to note here that the shift of $\tilde{\bf b}$
(the impact parameter of final pion)
is equivalent to the shift of the impact parameter of the initial proton according
to the relation between impact parameters in the multiparticle production\cite{webb}:
\begin{equation}\label{bi}
{\bf b}=\sum_i x_i{ \tilde{\bf  b}_i}.
\end{equation}
The variable $\tilde b$ is conjugated to the transverse momentum of $\pi^0$,
but relations  between functions depending on the impact parameters
$\tilde b_i$,  which will be used further for the calculation of asymmetry,
are nonlinear and therefore we are
using the semiclassical correspondence between small and large values
 of transverse momentum and impact parameter:
\begin{equation}\label{bp}
\mbox{small}\;\tilde b \Leftrightarrow \mbox{large}\;p_T \quad\mbox{and}\quad
\mbox{large}\;\tilde b \Leftrightarrow \mbox{small}\;p_T.
\end{equation}

We consider production of $\pi^0$ in the fragmentation region, i.e.
at large $x_F$ and therefore use the approximate relation
\begin{equation}\label{bx}
b\simeq x_F\tilde b,
\end{equation}
which results from Eq. (\ref{bi}) with an additional assumption on the
small values of Feynman $x_F$ for other particles. In the symmetrical case of $pp$-interactions
 the model assumes  equal mean multiplicities
in the forward and backward hemispheres and  small momentum transfer
between two sides. In that sense the model is similar the approach of
 Chou and Yang \cite{yang}.

 We apply chiral quark semiclassical mechanism which takes into account
unitarity in the direct channel
to obtain qualitative conclusions
on asymmetry dependence on the kinematical variables.

\section{Asymmetry  and inclusive cross-section}
  The main feature of the mechanism is an account of
unitarity in the direct channel of
reaction. The corresponding formulas for inclusive
cross--sections of the process
\[ h_1 +h_2^\uparrow\rightarrow h_3 +X, \] where hadron $h_3$ in this particular case
 is $\pi^0$ meson and $h_1$, $h_2$ are  protons,
 were obtained in
\cite{tmf} and have the following form
\begin{equation}
{d\sigma^{\uparrow,\downarrow}}/{d\xi}= 8\pi\int_0^\infty
bdb{I^{\uparrow,\downarrow}(s,b,\xi)}/ {|1-iU(s,b)|^2},\label{un}
\end{equation}
where $b$ is the impact  parameter of the initial
protons. Here the function
$U(s,b)$ is the generalized reaction matrix (averaged over initial spin states)
which is determined by the basic dynamics of elastic scattering.
 The elastic scattering amplitude in the impact
parameter representation $F(s,b)$
   is then given \cite{log}
 by the  relation:
  \begin{equation} F(s,b)=U(s,b)/[1-iU(s,b)].
\label{6} \end{equation}
This equation allows one to obey unitarity provided inequality
  $ \mbox{Im}\,U(s,b)\geq 0\,$  is fulfilled.
The functions $I^{\uparrow,\downarrow}$ in Eq. (\ref{un}) are related   to the
functions   $U_n^{\uparrow,\downarrow}$ --
  the multiparticle
analogs of the function $U$ \cite{tmf} in the polarized case.
The kinematical variables $\xi$
($x_F$ and $p_T$ for example) describe the state of the produced particle
$h_3$.
   Arrows $\uparrow$ and $\downarrow$ denote
   transverse spin directions of the polarized proton $h_2$.

Asymmetry  $A_N$
can be expressed in terms of the functions $I_{-}$, $I_{0}$ and $U$:
\begin{equation} A_N(s,\xi)=\frac{\int_0^\infty bdb
I_-(s,b,\xi)/|1-iU(s,b)|^2} {2\int_0^\infty bdb
I_0(s,b,\xi)/|1-iU(s,b)|^2},\label{xnn}
\end{equation}
where $I_0=1/2(I^\uparrow+I^\downarrow)$ and $I_-=(I^\uparrow-I^\downarrow)$
and $I_0$ obey the sum rule
\[
\int I_0(s,b,\xi) d\xi = \bar n(s,b)Im U(s,b),
\]
here $\bar n(s,b)$ stands for the mean multiplicity in the impact parameter
representation.

On the basis of the described mechanism we can
postulate that the functions
$I^\uparrow(s,b,\xi)$ and $I^\downarrow(s,b,\xi)$ are related to the functions
$\frac{1}{3}I_0(s,b,\xi)|_{\tilde b-\delta\tilde b }$ and $\frac{1}{3}I_0(s,b,\xi)|_{\tilde b+
\delta\tilde {b} }$,
respectively, i.e.
\begin{equation}\label{der}
I_-(s,b,\xi)=\frac{1}{3}[I_0(s,b,\xi)|_{\tilde {b}-\delta\tilde {b} }-
I_0(s,b,\xi)|_{\tilde{b}+\delta\tilde{b} }]
=-\frac{2}{3}\frac{\delta I_0(s,b,\xi)}{\delta\tilde{b}}\delta\tilde b.
\end{equation}

We can connect $\delta\tilde b$ with the radius of quark interaction
$r_{Q}^{flip}$
responsible for the transition  changing quark spin:
\[
\delta\tilde b\simeq r_{Q}^{flip}.
\]

Using the above relations and,
in particular, (\ref{bx}), we can write
the following expression for asymmetry $A_N^{\pi^0}$
\begin{equation} A_N^{\pi^0}(s,\xi)\simeq -x_Fr_{Q}^{flip}\frac{1}{3}\frac{\int_0^\infty bdb
I'_0(s,b,\xi)db/|1-iU(s,b)|^2} {\int_0^\infty bdb
I_0(s,b,\xi)/|1-iU(s,b)|^2},\label{poll}
\end{equation}
where $I'_0(s,b,\xi)={dI_0(s,b,\xi)}/{db}$.  In (\ref{poll})
we  made replacement according to relation (\ref{bx}):
\[
{\delta I_0(s,b,\xi)}/{\delta\tilde{b}}\Rightarrow {dI_0(s,b,\xi)}/{db}.
\]
It is clear that $A_N^{\pi^0}(s,\xi)$
(\ref{poll})
should be positive because $I'_0(s,b,\xi)<0$.

The function $U(s,b)$ is
chosen  as a product of the averaged quark amplitudes
in accordance with the quasi-independence of valence constituent
quark scattering in the self-consistent mean field \cite{csn}. The generalized
reaction matrix $U(s,b)$ (in a pure imaginary case, which we consider
 here for simplicity) is
the following
\begin{equation} U(s,b) = i\tilde U(s,b)=ig(s)\exp(-Mb/\zeta ),
 \label{x}
\end{equation}
where the function $g(s)$ power like increases at large values of $s$
\[
g(s)\sim  \left (\frac{\sqrt{s}}{m_Q}\right)^N,
\]
$M$ is the total mass of $N$ constituent quarks with mass $m_Q$ in
the initial hadrons and parameter $\zeta$  determines  a universal scale for
the quark interaction radius in the model, i.e. $r_Q=\zeta /m_Q$.

To evaluate asymmetry dependence on $x_F$ and $p_T$
we use semiclassical correspondence  between transverse momentum and impact parameter
  values, (\ref{bp}).
Performing integration by parts and
choosing the region of  small $p_T$ we  select  the large values of impact parameter
 and we obtain
\begin{equation} A_N^{\pi^0}(s,\xi)\sim x_Fr_{Q}^{flip}
\frac{M}{3\zeta}\frac{\int_{b>R(s)} bdb
I_0(s,b,\xi)\tilde U(s,b) } {\int_{b>R(s)} bdb
I_0(s,b,\xi)},\label{pollsm}
\end{equation}
where $R(s)\sim \ln s$ is the hadron interaction radius, which serve as a scale
separating large and small impact parameter regions.
In the large  impact parameter region:
$\tilde U(s,b)\ll 1$ for $b\gg R(s)$ and therefore
 we have a small dynamically suppressed asymmetry $ A_N^{\pi^0}(s,\xi)$
in the region of small and moderate values of $p_T$, i.e. $p_T\leq x_F/R(s)$.

But at small values of $b$  the value of $U$-matrix is large,
 $\tilde U(s,b)\gg 1$,
 and  we can neglect unity in the denominators of the integrands.
\begin{figure}[htb]
\begin{center}
  \resizebox{6.5cm}{!}{\includegraphics*{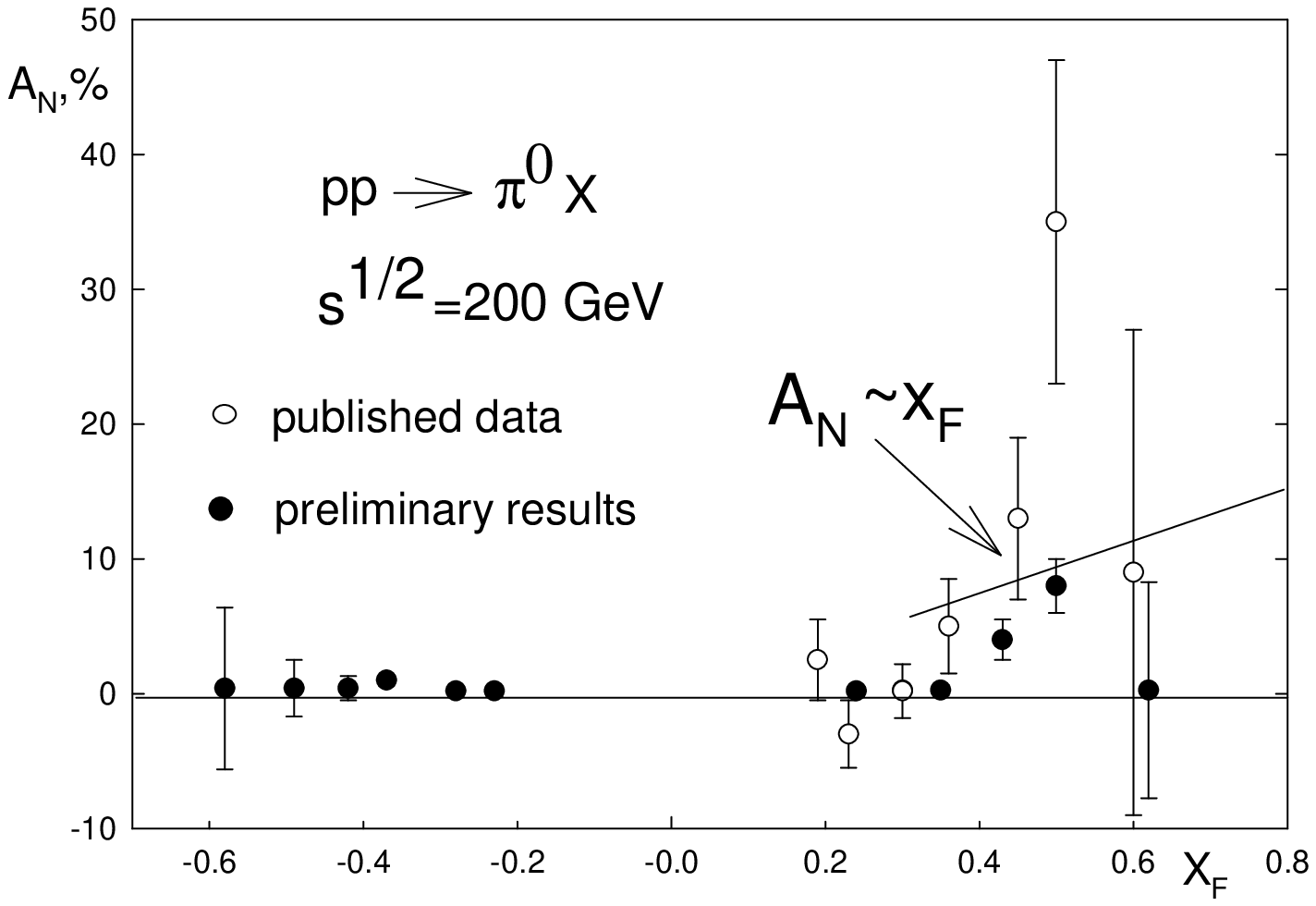}}\;\;\quad
  \resizebox{6.5cm}{!}{\includegraphics*{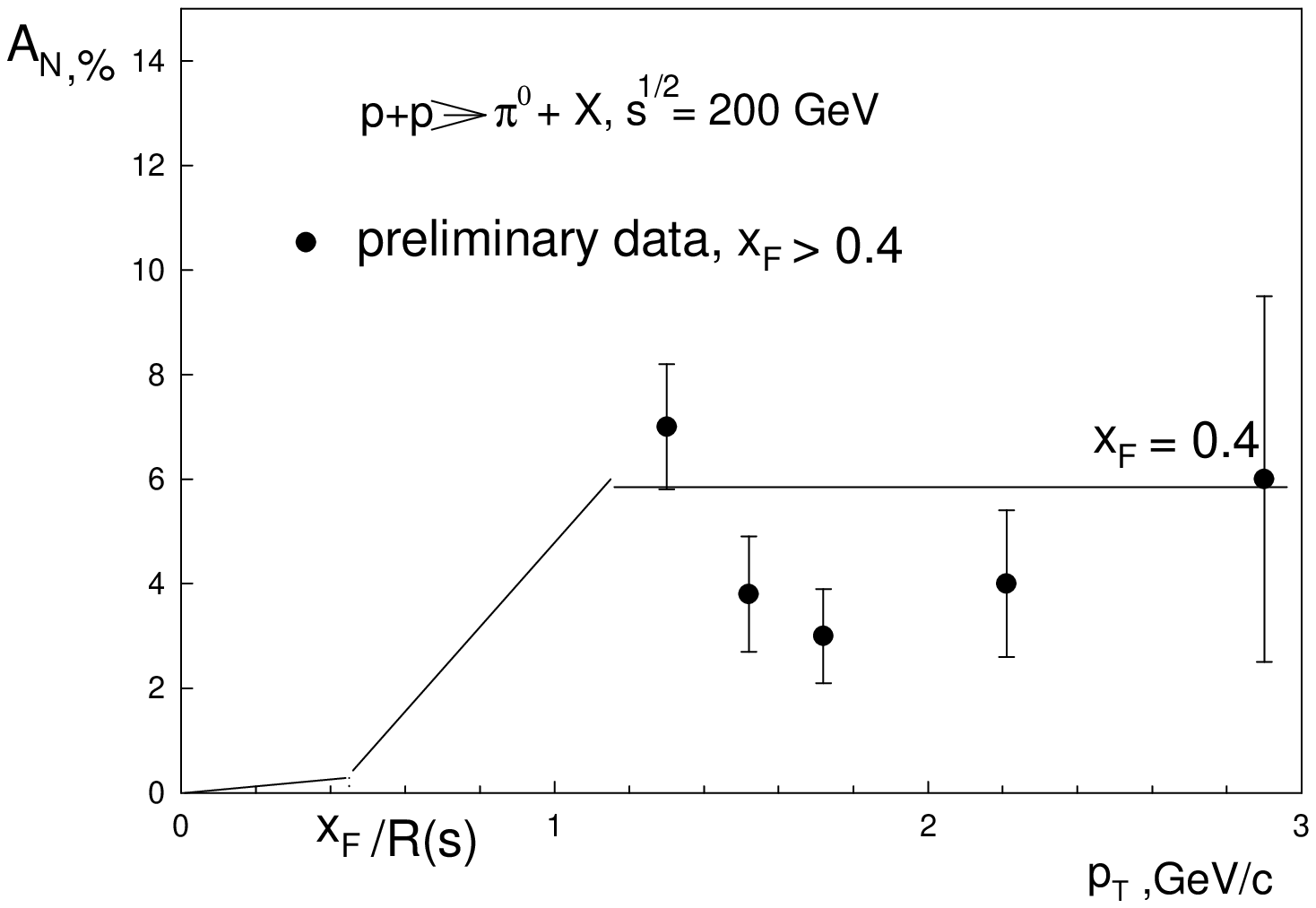}}
\end{center}
\caption{$x_F$ (left panel) and $p_T$ (right panel)
 dependencies of the asymmetry $A_N$ in the process $p+p_\uparrow\to\pi^0+X$ at RHIC,
  experimental data from \cite{star}.} \label{ts}
\end{figure}
Thus the ratio of two integrals (after integration by parts of nominator in Eq. (\ref{poll}))
 is of order
of unity, i.e.  the energy and $p_T$-independent behavior
of asymmetry $A_N^{\pi^0}$ takes place at the values of transverse momentum $p_T\gg x_F/R(s)$:
\begin{equation} A_N^{\pi^0}(s,\xi)\sim x_Fr_{Q}^{flip}
\frac{M}{3\zeta}.\label{polllg}
\end{equation}
This flat transverse momentum dependence of asymmetry results from the similar
rescattering effects for the different spin states, i.e. spin-flip and spin-nonflip
interactions undergo similar absorption at short distances and
the relative magnitude of this absorption does not depend on energy. It is one
of the manifestations of the unitarity.
The numeric value of polarization $A_N^{\pi^0}$ can be significant; there are
no small factors in (\ref{polllg}). In Eq. (\ref{polllg}) $M$ is
proportional to two nucleon masses, the value of parameter $\zeta \simeq 2$.
 We expect that $r_{Q}^{flip}\simeq
0.1-0.2$ fm on the basis of the model estimate \cite{csn,tmf}.  The above qualitative
 features of asymmetry dependence on $x_F$,
$p_T$ and energy are in a good agreement with the experimentally observed trends
 \cite{star}.
For example, Fig. 2 demonstrates that the linear $x_F$ dependence is in a good agreement with
the experimental data of STAR Collaboration at RHIC \cite{star}
 in the fragmentation region ($x_F\geq 0.4$) where the model
should be applicable. Of course,
the conclusion on the $p_T$--independence of polarization is a  qualitative one
and small deviations from such behavior cannot be excluded.
The same dependencies are compared with the FNAL E704 data \cite{e704} (Fig.3).
Those dependencies as it is clear from their derivations are valid in high-energy
approximation and therefore have been compared with FNAL and RHIC data only. However,
they are in qualitative agreement with the lower energy data also \cite{prz}.
\begin{figure}[htb]
\begin{center}
  \resizebox{6.5cm}{!}{\includegraphics*{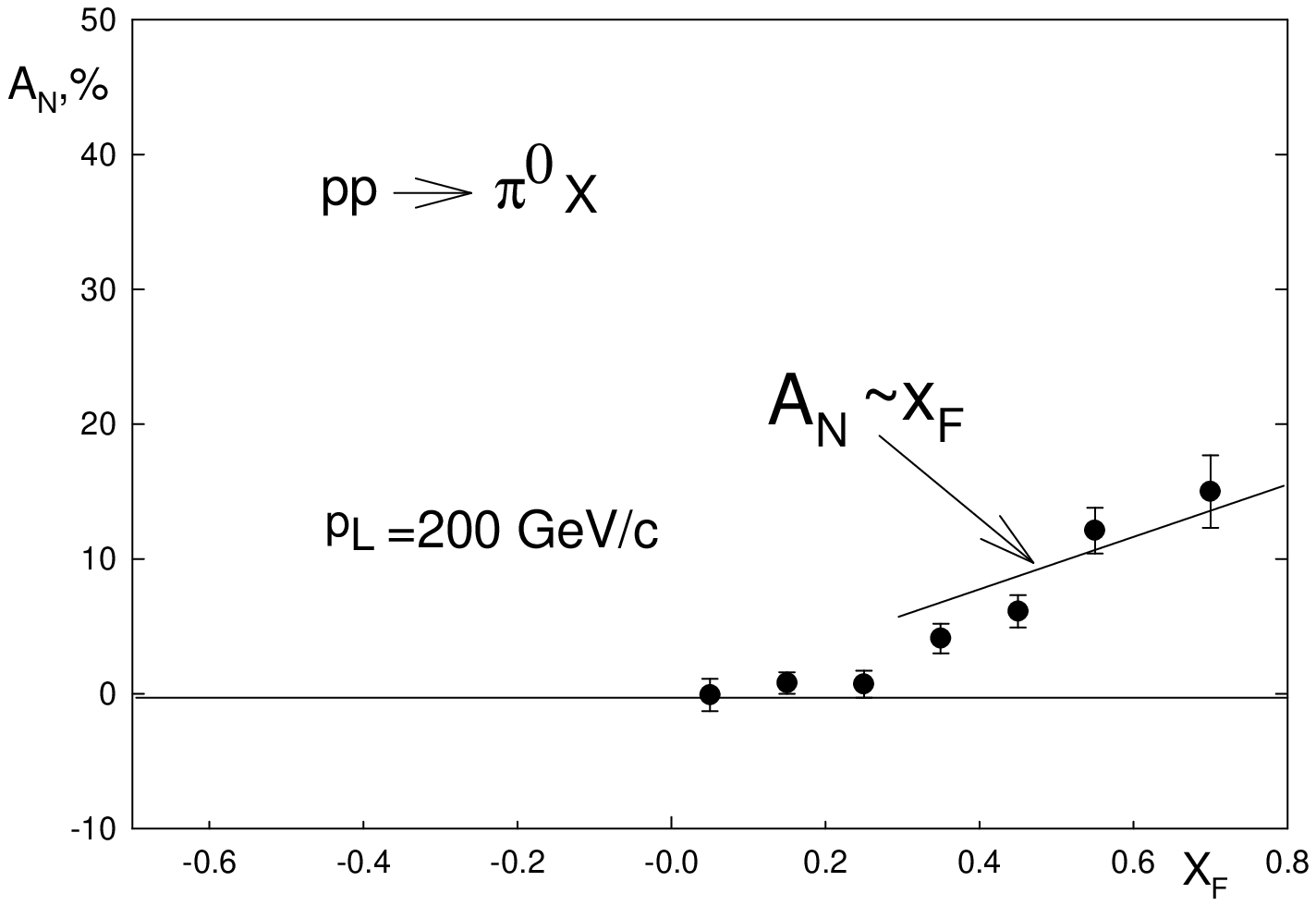}}\;\;\quad
  \resizebox{6.5cm}{!}{\includegraphics*{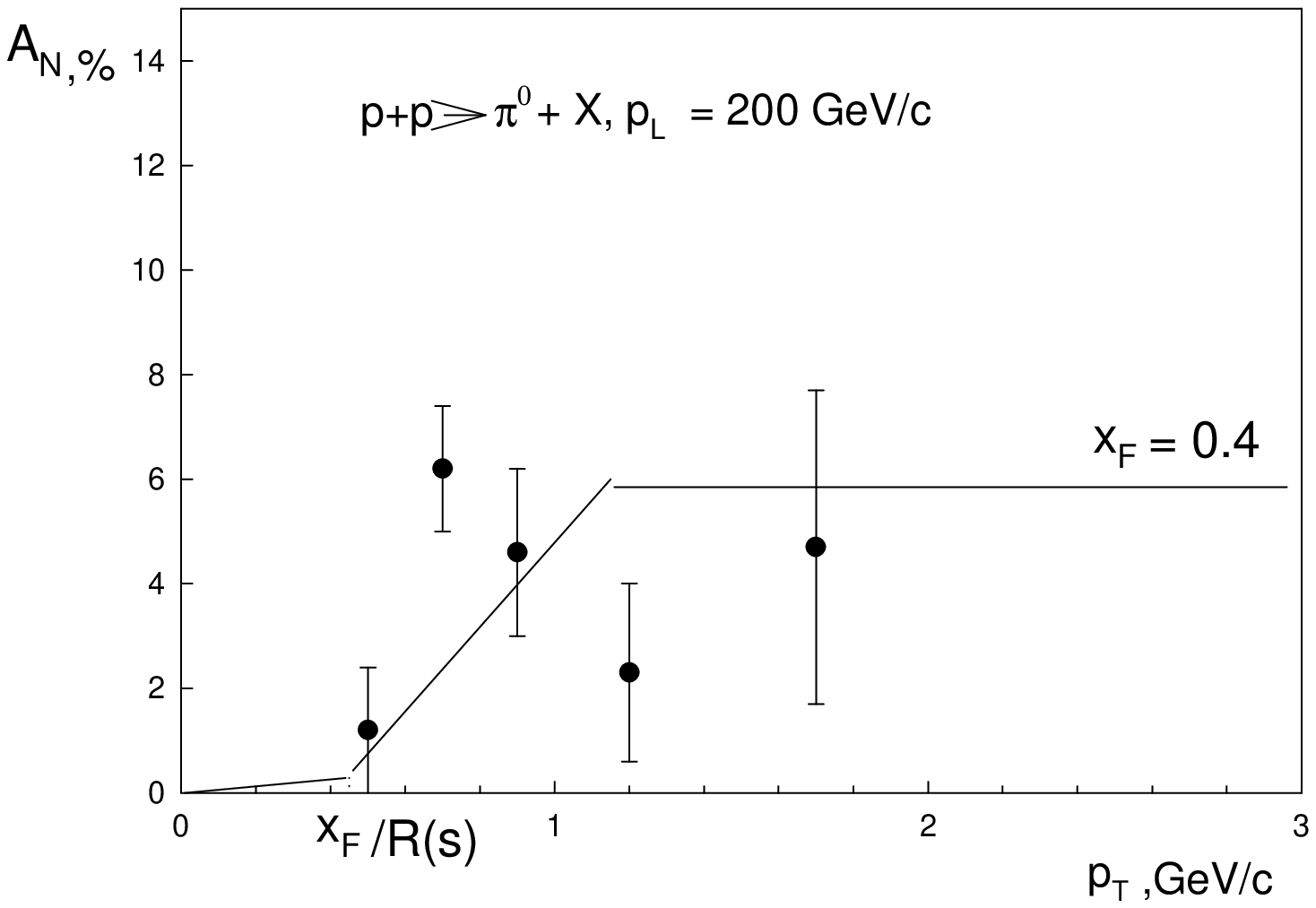}}
\end{center}
\caption{$x_F$ (left panel) and $p_T$ (right panel)
 dependencies of the asymmetry $A_N$ in the process $p+p_\uparrow\to\pi^0+X$ at FNAL,
  experimental data from \cite{e704}.} \label{tsf}
\end{figure}
Similar mechanism should generate SSA in the production of charged pions.
The relevant process for $\pi^+$--production in polarized $pp_\uparrow$ interactions
\[
U_\uparrow\to\pi^+ + D_\downarrow,
\]
leads to a negative shift in the impact parameter and consequently
to the positive asymmetry $A_N$, while the corresponding process for
the $\pi^-$--production
\[
D_\downarrow\to\pi^- + U_\uparrow
\]
 leads to the positive shift in impact parameter and, respectively,
 to the negative asymmetry $A_N$.
Asymmetry $A_N$ in the $\pi^{\pm}$-production in the
 fragmentation region of polarized proton should have linear $x_F$--dependence
  at $x_F>0.4$ and flat $p_T$ dependence at large $p_T $. Those dependencies
  are similar to the ones depicted on Fig. 2 for $\pi^0$--production.

The reversed mechanism (chiral quark spin filtering) was used for
the explanation of the hyperon polarization \cite{hypern}. Note that
polarization of $\Lambda$ -- hyperon has the same generic
 dependence on $x_F$ and $p_T$.

To demonstrate the model self-consistency it should be noted that it is
able to describe
the  unpolarized cross-section of $\pi^0$-production also.
In the fragmentation region it was shown \cite{unpol} that
at small $p_T$ the poles in impact parameter plane at $b\sim R(s)$
lead to the exponential $p_T$--dependence of inclusive cross-section.
\begin{figure}[htb]
\begin{center}
  \resizebox{8cm}{!}{\includegraphics*{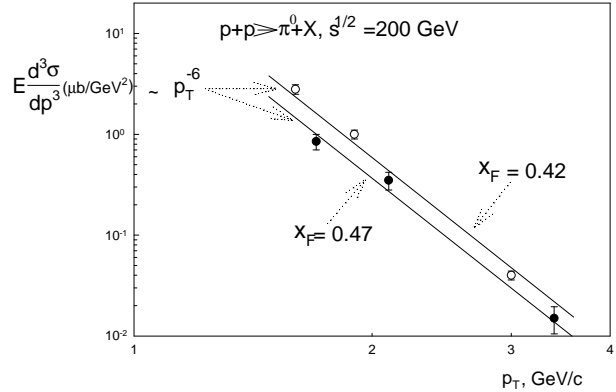}}
\end{center}
\caption{Transverse momentum dependence of unpolarized inclusive cross--section,
experimental data from \cite{star}.} \label{ds}
\end{figure}
At high $p_T$ the power-like dependence $p_T^{-n}$ with $n=6$
should take place. It originates from the singularity at zero impact parameter
 $b=0$.
 The exponent $n$ does not depend
 on $x_F$. Experimental data are in a good agreement with the $p_T^{-6}$--dependence of
the unpolarized inclusive cross--section (Fig. 3). Recently a similar $p_T^{-6}$--dependence
has been obtained for the  soft contribution to quark-quark scattering induced by an anomalous
 chromomagnetic interaction due to instanton mechanism \cite{kpt6}.

Thus, in the approach with effective degrees of freedom --
 constituent quarks and Goldstone bosons -- differential cross--section at high transverse momenta
 has a generic power-like dependencies on $p_T$
 at large transverse momenta.  The differential cross-section and the asymmetry $A_N$
   are in  agreement with the experimental data at the highest available energy at RHIC and
asymmetry is in agreement with FNAL data also.

\section*{Conclusion}
The proposed semiclassical mechanism of SSA generation considers the
 effective degrees of freedom and takes into
account collective aspects of QCD dynamics. Together with unitarity, which is an essential
part of this approach, it leads to
 linear dependence on $x_F $ and
flat dependence of SSA on transverse momentum at large $p_T$
in the polarized proton fragmentation region.
Such dependencies  with the energy independent
behavior of asymmetry at large transverse
momenta are the direct phenomenological
consequences of the proposed mechanism.

The chiral quark fluctuation  mechanism in effective field with spin flip is
 relatively suppressed when compared to direct elastic
 scattering of quarks in effective   field and therefore
does not play a role in the reaction $pp_\uparrow\to pX$ in the fragmentation
 region, but it should not be suppressed in $pp_\uparrow\to nX$. These features really
  take place in the experimental data set: asymmetry $A_N$ is zero for proton production
 and deviates from zero for neutron production in the forward region.

We discussed here  particle production in the fragmentation region of polarized proton.
In the symmetrical case of $pp$-interactions the model leads to the equal mean multiplicities
in the forward and backward hemispheres and very small momentum transfer is expected
between the two sides.
In the central and backward regions where correlations
 between impact parameter of the initial and impact parameters of the final particles
 are weak or completely degraded, the asymmetry cannot be generated
 due to this chiral quark semiclassical mechanism. The vanishing asymmetries
 in the central and backward regions observed experimentally provide the indirect
evidences  in favor of this mechanism.
\section*{Acknowledgement}
We are grateful to S. Shimanskiy and A. Vasiliev for the
interesting discussions.

\end{document}